\documentclass[conference]{IEEEtran}
\IEEEoverridecommandlockouts
\usepackage{cite}
\usepackage{amsmath,amssymb,amsfonts}
\usepackage{algorithmic}
\usepackage{graphicx}
\usepackage{textcomp}
\usepackage{xcolor}
\usepackage{subcaption}
\usepackage{url}
\usepackage{multirow}

\newcommand{\yuhuarevised}[1]{{\color{black}#1}}
\newcommand{\TKrevised}[1]{{\color{black}#1}}
\newcommand{\Kevinrevised}[1]{{\color{black}#1}}

\def\BibTeX{{\rm B\kern-.05em{\sc i\kern-.025em b}\kern-.08em
    T\kern-.1667em\lower.7ex\hbox{E}\kern-.125emX}}
\author{Author Name}
\begin{document}

\title{Explicit Note-Event Tokenization and Pitch-Validity Constrained Decoding for MIDI-to-Tablature Transcription}


\author{
\begin{tabular}{cc}
\begin{minipage}[t]{0.45\textwidth}
\centering
1\textsuperscript{st} Ting-Kai Hsu\\
\textit{Graduate Institute of Communication Engineering}\\
\textit{National Taiwan University}\\
Taipei, Taiwan\\
r13942132@ntu.edu.tw
\end{minipage}
&
\begin{minipage}[t]{0.45\textwidth}
\centering
2\textsuperscript{nd} Wei-Chin Wang\\
\textit{Department of Electrical Engineering}\\
\textit{National Taiwan University}\\
Taipei, Taiwan\\
b10502010@ntu.edu.tw
\end{minipage}
\\
\noalign{\vskip 2.5em}
\begin{minipage}[t]{0.45\textwidth}
\centering
3\textsuperscript{rd} Kai-Xi Hong\\
\textit{Department of Electrical Engineering}\\
\textit{National Taiwan University}\\
Taipei, Taiwan\\
b10401006@ntu.edu.tw
\end{minipage}
&
\begin{minipage}[t]{0.45\textwidth}
\centering
4\textsuperscript{th} Yu-Hua Chen\\
\textit{Graduate Institute of Networking and Multimedia}\\
\textit{National Taiwan University}\\
Taipei, Taiwan\\
f08946011@ntu.edu.tw
\end{minipage}
\end{tabular}
}

\maketitle



\begin{abstract}

Guitar tablature transcription predicts the string and fret position for each note so that the resulting tablature reproduces the target musical part. Prior sequence-to-sequence approaches have shown promising results on large-scale datasets, but their generalization behavior across different dataset scales remains less explored. In this work, we propose a guitar tablature transcription framework with explicit note-event tokenization and regularized training. The proposed decoder \TKrevised{token} representation incorporates note-event tokens together with TAB tokens, allowing note boundaries, pitch-related events, and string--fret positions to be represented more explicitly.

We evaluate the proposed framework on DadaGP, a large-scale dataset, and Fran\c{c}ois Leduc, a small-scale dataset. Our method improves tablature accuracy over the Fretting Transformer baseline on \TKrevised{DadaGP}, with especially strong gains when trained directly on the small-scale Leduc dataset. We further introduce a pitch-validity constrained decoding strategy that masks \yuhuarevised{pitch-}invalid TAB candidates during generation rather than correcting them after decoding \TKrevised{and simultaneously preserves the original timing and note structure from the input}. This constraint improves tablature accuracy and provides a controlled setting for measuring how much error remains after pitch-invalid predictions are removed. 
Our code will be released at: \url{https://github.com/MusicGuitarTab/GuitarTab}

\end{abstract}


\section{Introduction}\label{sec:introduction}


Guitar tablature (TAB) is a form of music notation that explicitly specifies the string and fret positions for each note on the guitar fretboard. 
By contrast, Musical Instrument Digital Interface (MIDI) is widely used as a symbolic music representation that encodes musical events, such as pitch, timing, and control information~\cite{MMT-BERT}, but it does not specify the physical playing position of each note on the guitar.
Therefore, symbolic guitar tablature
transcription is not merely a pitch transcription task. Instead, it is an inference problem in which each pitch event must be mapped to one of
several possible string--fret realizations. Classical approaches have addressed this ambiguity using explicit optimization methods, such as dynamic programming and genetic algorithms~\cite{sayegh1989fingering,tuohy2005ga,ramos2015comparative}.

Recent transformer-based approaches formulate \TKrevised{guitar tablature} transcription as a sequence modeling problem, learning the mapping from MIDI-like event representations to TAB tokens~\cite{fretting-transformer,midi-to-tab}.
These methods have demonstrated promising results on large-scale tablature datasets, such as DadaGP~\cite{dadagp}. However, two important aspects remain insufficiently explored. 
First, the role of target-side tokenization in tablature generation has received limited attention. In autoregressive formulations such as the Fretting Transformer~\cite{fretting-transformer}, the decoder output mainly consists of \texttt{TIME\_SHIFT\_TICKS} and \texttt{TAB<string,fret>} tokens. This representation explicitly encodes the physical string--fret position, but note boundaries and pitch-related events are only implicit in the generated sequence. 
Second, prior work often relies on large-scale datasets, synthetic data, or combined-dataset training, leaving the effect of dataset scale on model generalization less clear.

Motivated by these observations, we propose an explicit note-event tokenization for guitar tablature transcription from MIDI-like event representations. 
Instead of generating only timing and TAB tokens, our decoder generates a structured sequence containing 
\TKrevised{\texttt{NOTE\_ON\Kevinrevised{\_PITCH}}, \texttt{TAB<string,fret>}, and \texttt{NOTE\_OFF\Kevinrevised{\_PITCH}} tokens and \texttt{TIME\_SHIFT\Kevinrevised{\_TICKS}}.}
The \texttt{NOTE\_ON\TKrevised{\_PITCH}} and \texttt{NOTE\_OFF\TKrevised{\_PITCH}} tokens expose note boundaries and pitch-related information on the target side, while the \texttt{TAB<string,fret>} token specifies the physical realization of each note.
This design aligns the target-side representation more closely with the MIDI-like input event structure and separates note-event modeling from string--fret assignment.

\TKrevised{In addtion to} decoder \TKrevised{token representation}, we introduce a pitch-validity constrained decoding strategy for guitar tablature generation. 
Since the input MIDI-like event sequence provides the pitch of each target note, the decoder can restrict TAB candidates to string--fret positions that physically realize the corresponding pitch. 
This constraint prevents pitch-invalid TAB tokens from being selected during decoding and provides a diagnostic setting for separating pitch consistency from the remaining ambiguity of choosing among multiple pitch-valid string--fret positions.

We evaluate the proposed framework under two dataset-scale regimes: DadaGP~\cite{dadagp}, a large-scale tablature dataset, and the Fran\c{c}ois Leduc dataset~\cite{leduc}, a small-scale dataset. 
For fair comparison, we retrain the Fretting Transformer separately on each dataset and compare it with our method under the same dataset-specific setting. 
In our experiments, the baseline can be trained successfully on DadaGP but tends to overfit on the Fran\c{c}ois Leduc dataset, whereas our full framework achieves better pitch and tablature accuracy in both settings.

In addition to standard transcription metrics, we conduct note-level error analysis and pitch-validity constrained decoding experiments to better understand the remaining challenges of guitar tablature transcription. In the constrained setting, pitch-invalid tablature candidates are removed during decoding using the target pitch, \yuhuarevised{while the timing and note structure are preserved from the input.} \TKrevised{This} analysis estimates the upper-bound performance when pitch-level errors are controlled, and also \Kevinrevised{demonstrates} the improvement of \Kevinrevised{token and tablature accuracy} after applying pitch-validity constrained decoding in both models.
Our contributions are summarized as follows:
\begin{itemize}
    \item We propose an explicit note-event decoder \TKrevised{token representation} for guitar
    tablature transcription from MIDI-like event representations. The proposed tokenization exposes note boundaries and pitch-related events on the target side while preserving string--fret tablature information.

    \item Experimental results show that our full framework improves pitch and tablature accuracy over the retrained Fretting Transformer baseline on DadaGP dataset.
    Further analysis shows that pitch-valid string--fret ambiguity remains a major source of error.

    \item We introduce a pitch-validity constrained decoding strategy for guitar tablature generation. By masking string--fret candidates that are inconsistent with the source MIDI pitch, the proposed decoding strategy reduces pitch-invalid TAB predictions and provides a diagnostic setting for analyzing string--fret ambiguity.
\end{itemize}

\section{Related Work}

\subsection{Symbolic Sequence Modeling and Token Representation}

Symbolic music modeling commonly represents musical content as discrete token sequences, where the design of event tokens determines how temporal, pitch, and structural information is exposed to the model. 
Self-attention-based models, such as Music Transformer and its variants, have demonstrated strong ability to capture long-range musical structure from symbolic token sequences~\cite{huang2019musictransformer, huang2020pop, shih2022theme, hsiao2021compound,chen2020automaticcompositionguitartab}.

For guitar tablature transcription, token representation is also critical because it determines how musical timing, pitch-related note events, and the corresponding string and fret information are encoded in the sequence.
This representation affects how explicitly the model can learn the relationship between pitch events and their possible playing positions.
Recent studies have adopted transformer-based models for symbolic guitar tablature transcription. 
MIDI-to-Tab~\cite{midi-to-tab} uses a BART-based~\cite{bart} encoder--decoder Transformer with a masked language modeling objective, where string tokens are masked and predicted from MIDI-derived symbolic tokens. 
Since pitch information is given in the input, the predicted string can be used to derive the corresponding fret, making the task closer to sequence labeling.
Fretting Transformer~\cite{fretting-transformer} formulates \TKrevised{guitar tablature} as an autoregressive sequence-to-sequence generation problem, where the decoder generates TAB tokens.
However, the decoder output format in the Fretting Transformer consists of \texttt{TIME\_SHIFT\TKrevised{\_TICKS}} and \texttt{TAB<string,fret>} tokens, which provides limited explicit note-event information during generation.

\subsection{Fingering Ambiguity and Playability}

Symbolic tablature inference is inherently ambiguous because a single pitch can
often be played at multiple string--fret positions. For example, in standard
tuning, A2 can be played either on the sixth string at the fifth fret or on the
open fifth string. Thus, a MIDI pitch sequence does not uniquely determine a
tablature sequence.

Classical approaches address this ambiguity through constrained optimization.
Dynamic programming, genetic algorithms, and probabilistic models have been used
to search for playable fingering sequences under physical or playability-related
constraints~\cite{sayegh1989fingering, tuohy2005ga, ramos2015comparative,
barbancho2009hmmguitar, hori2013iohmm, Ariga2017Song2GuitarAD}. These methods
explicitly model feasibility and playability, but they often rely on manually
designed cost functions and local transition assumptions.

In contrast, Transformer-based models~\cite{fretting-transformer} can learn string--fret preferences from
data, but they may still produce pitch-invalid or dataset-inconsistent TAB
predictions. 

\subsection{Constrained Decoding for Structured Generation}
Constrained decoding restricts the set of valid output tokens during generation
without retraining the model. It has been used in structured generation tasks
such as lexically constrained translation and text-to-SQL generation, where
invalid tokens are filtered during decoding~\cite{
hokamp2017lexicallyconstraineddecodingsequence,
post2018fastlexicallyconstraineddecoding, scholak2021picard}.

In \TKrevised{guitar tablature}, pitch-validity \Kevinrevised{constraint has} also been used as post-generation correction. \Kevinrevised{The Fretting Transformer~\cite{fretting-transformer} applies rule-based post-processing techniques, including overlap correction and neighborhood search, to correct generated string--fret combinations that produce incorrect pitches.}

\begin{figure*}[t]
    \centering
    \includegraphics[width=0.85\textwidth]{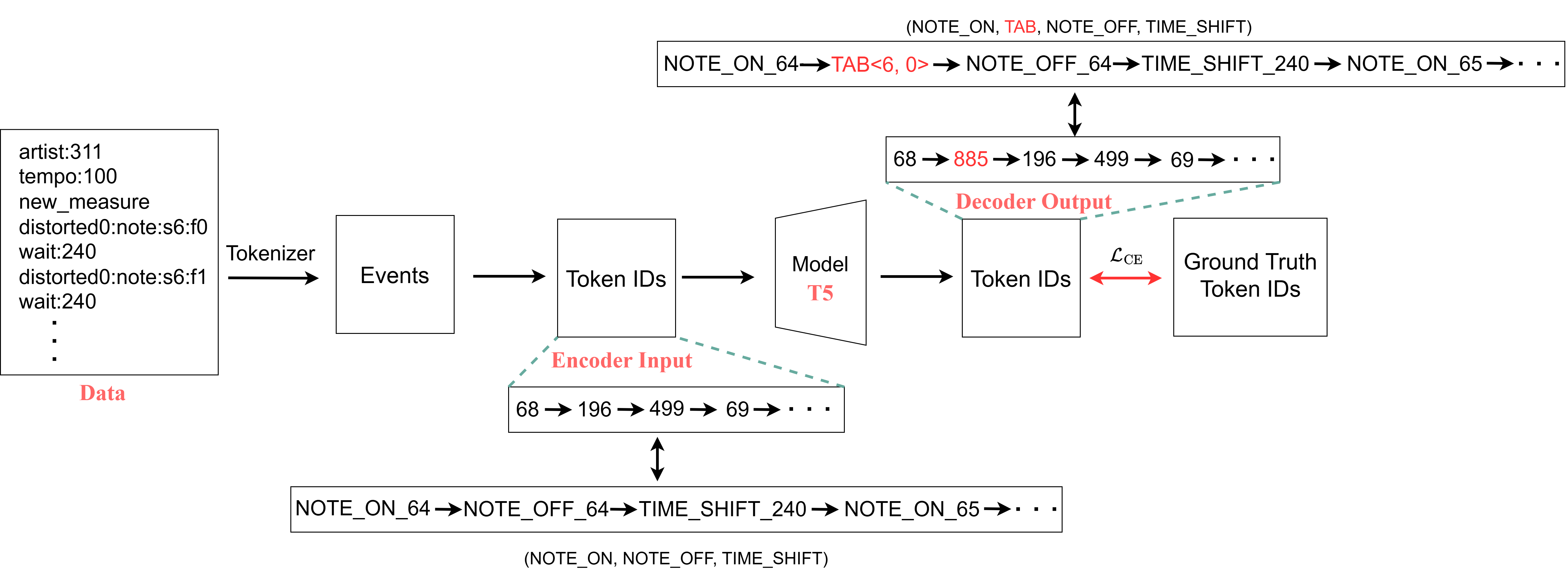}
    \caption{Overview of the proposed framework for guitar tablature transcription. The model takes MIDI-like event tokens as input and generates a token sequence that includes TAB tokens for representing string and fret positions.}
    \label{fig:framework}
\end{figure*}


\section{Methodology}\label{sec:methodology}
We formulate guitar tablature transcription as a sequence-to-sequence translation problem from MIDI-derived note events to a structured
token sequence.
As shown in Figure~\ref{fig:framework}, the raw symbolic data are tokenized into event tokens and converted into token IDs before being fed into the T5 encoder--decoder model. The input \TKrevised{token }sequence contains \texttt{NOTE\_ON\_PITCH}, \texttt{NOTE\_OFF\_PITCH}, and \texttt{TIME\_SHIFT\_TICKS} tokens.
\texttt{TIME\_SHIFT\_TICKS} represents a discrete time interval \TKrevised{measured in ticks, where one quarter note equals 960 ticks.}
The output \TKrevised{token }sequence specifies both the note-event structure and the physical string--fret realization of each note. The detailed decoder token \TKrevised{representation} is described below.


\subsubsection{Decoder Token Representation}

Unlike the Fretting Transformer, whose decoder vocabulary mainly consists of
\texttt{TIME\_SHIFT\_TICKS} and \texttt{TAB<string,fret>} tokens, our
decoder also includes \texttt{NOTE\_ON\_PITCH} and
\texttt{NOTE\_OFF\_PITCH} tokens. For each note event, the decoder generates
a \texttt{NOTE\_ON\_PITCH} token together with a
\texttt{TAB<string,fret>} token that specifies its playing position, while
\texttt{TIME\_SHIFT\_TICKS} and \texttt{NOTE\_OFF\_PITCH} tokens encode
timing and note-release events.

\begin{figure}[h]
    \centering
    \includegraphics[width=0.9\linewidth]{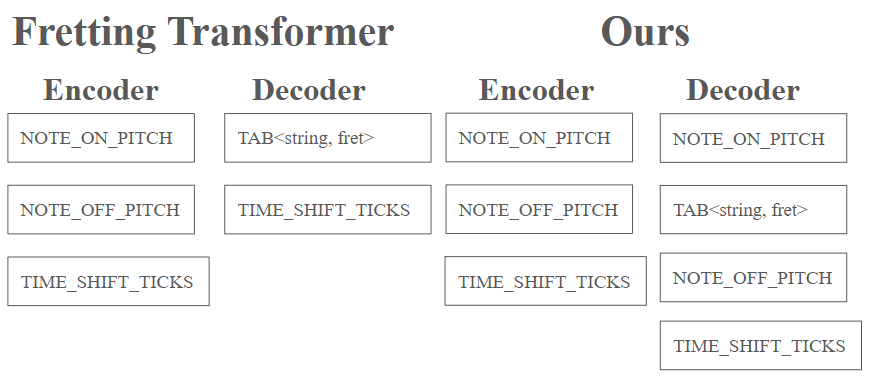}
    \caption{\TKrevised{Comparison of the decoder token representations used by the Fretting Transformer and our method.}}
    \label{fig:token_scheme}
\end{figure}

By adding explicit note-event tokens, the \TKrevised{output} sequence exposes note
boundaries and pitch-related information during generation. Figure~\ref{fig:token_scheme}
compares the proposed decoder token representation with the Fretting Transformer.

\subsubsection{Architecture}
We use T5~\cite{T5}, a Text-to-Text Transfer Transformer, as our backbone model.
The encoder consumes the input event tokens and produces contextual representations, and the decoder autoregressively generates the target TAB token sequence.
Our configuration is: $d_{\text{model}}=128$, $d_{\text{ff}}=1024$, $N_{\text{enc}}=3$, $N_{\text{dec}}=3$, and the number of heads is $h=4$. For regularization, we use a weight decay coefficient of $0.1$ and a dropout rate of $0.1$.

\subsubsection{Training objective}
We train the model with teacher forcing to minimize the token-level cross-entropy between the predicted sequence and the ground-truth sequence:
\begin{equation}
    L_{\text{CE}} = -\sum_{i=1}^{n} \log p\left(y_i \mid y_{<i}, x\right),
\end{equation}
where $x$ is the input event-token sequence, $y_i$ is the ground-truth token at position $i$, and $p(\cdot)$ is the decoder's predicted distribution.

\subsubsection{Pitch-Validity Constrained Decoding}\label{sec:constrained_decoding}

The Fretting Transformer applies a rule-based post-processing step to correct invalid predictions after the complete output \TKrevised{token }sequence has been generated~\cite{fretting-transformer}.
In contrast, we apply the pitch-validity constraint during decoding. This prevents pitch-invalid \texttt{TAB<string,fret>} tokens from being selected in the autoregressive generation process \TKrevised{and simultaneously preserves the original timing and note structure from the input, so that only the string--fret ambiguity remains to be resolved}.

Since the input MIDI-like sequence specifies the pitch of each note event, we use the source pitch at each TAB prediction step to construct a valid set of string--fret candidates. For a target pitch $p$, the pitch-valid TAB set is defined as
\begin{equation}
    \mathcal{T}(p) = \bigl\{(s,\,f) \;\big|\; \text{tuning}[s] + f = p,\;
    s\!\in\![1,6],\; f\!\in\![0,24] \bigr\},
\end{equation}


where $\text{tuning}[s]$ is the MIDI pitch of the open string $s$, and $f=0$ denotes an open
string. We index strings from the lowest-pitched to the highest-pitched
string, so the standard-tuning open-string pitches are
$o=[40,45,50,55,59,64]$, corresponding to E2, A2, D3, G3, B3, and E4.
At each decoding step corresponding to a \texttt{TAB<string,fret>} token, all logits associated with string--fret pairs outside $\mathcal{T}(p)$ are set to $-\infty$ before token selection. 

By construction, all generated TAB tokens in this setting are pitch-valid with respect to the source pitch skeleton. Therefore, pitch-validity constrained decoding is used as a diagnostic setting to estimate the performance achievable when pitch-invalid TAB candidates are removed and to analyze the remaining difficulty of selecting the ground-truth string--fret position among multiple pitch-valid candidates.

\section{Experiment}\label{sec:experiment}

\subsection{Dataset}
We select 5,185 songs from the DadaGP dataset \cite{dadagp}. For consistency, we retain only a single guitar track per song and restrict the data to 6-string guitar pieces, stored as GuitarPro \texttt{.gp} files. We split the dataset into training, validation, and test sets with 4,148 / 518 / 519 songs, respectively. 


For the Fran\c{c}ois Leduc dataset \cite{leduc}, we use 79 songs in total, split into training, validation, and test sets with 55 / 16 / 8 songs, respectively. Due to the limited size of the training set, we apply pitch transposition augmentation by shifting each training song by $\pm$1 to $\pm$5 semitones. Finally, we get 605 training files (55 songs $\times$ 11 transpositions).

\subsection{Evaluation Metrics}

We evaluate transcription correctness using three metrics. (i) \textit{Token Accuracy} measures the \TKrevised{token-level exact-match rate between the predicted sequence and the ground-truth target sequence}, reflecting overall generation quality. (ii) \textit{Pitch Accuracy} measures whether the pitch inferred from each TAB token matches the pitch inferred from the corresponding ground-truth TAB token. (iii) \textit{Tablature Accuracy} measures the match  of the predicted string and fret positions to the ground truth at the corresponding TAB tokens. 

Furthermore, We additionally report \textit{Difficulty Score}  \cite{fretting-transformer} based on fretboard movements and fingering transitions as a reference metric for 
playability, where lower scores indicate more playable tablature.

\subsection{Results}
Tables~\ref{tab:dadagp_train_setting} and~\ref{tab:leduc_train_setting} present the evaluation results on the DadaGP and Leduc test sets under different training data settings. 
All models are trained for 300 epochs in each setting. 
Bold indicates the better result within each training setting.
As shown in Table~\ref{tab:dadagp_train_setting}, our method outperforms the Fretting Transformer across all transcription metrics, except for difficulty metric. 
When trained only on DadaGP, our method achieves better token, pitch, and \Kevinrevised{tablature accuracy} than the Fretting Transformer, with improvements of \Kevinrevised{10.56\%, 4.4\%, and 5.32\%}, respectively.
This result suggests that the proposed method helps the model more effectively learn the mapping from MIDI-like event representations to TAB tokens.
Under the combined training setting, our method shows a slight improvement across all transcription metrics, whereas the Fretting Transformer shows a small accuracy drop.

We further evaluate both methods on the Leduc test set, as shown in Table~\ref{tab:leduc_train_setting}. When trained only on Leduc, the Fretting Transformer severely overfits and achieves only 0.49\% tablature accuracy, even after hyperparameter tuning. 
\Kevinrevised{In contrast, our method demonstrates stable training on the
small-scale Leduc dataset and achieves 98.06\% tablature
accuracy.}
Under the combined training setting, the Fretting Transformer obtains slightly higher accuracy than our method on the Leduc test set. 

Nevertheless, considering the limited scale of Leduc, our method remains competitive in the combined setting and demonstrates better stability when trained directly on the small-scale dataset.



\begin{table}[t]
\centering
\renewcommand{\arraystretch}{1.15}
\resizebox{\columnwidth}{!}{
\begin{tabular}{l|cccc}
\hline
\textbf{Method} & \textbf{Token Acc.$\uparrow$} & \textbf{Pitch Acc.$\uparrow$} & \textbf{Tab Acc.$\uparrow$} & \textbf{Difficulty $\downarrow$} \\
\hline
\multicolumn{5}{c}{\textit{Test on DadaGP (trained on DadaGP only)}} \\
\hline
Fretting Transformer & 80.57\% & 85.66\% & 71.78\% & \textbf{3.138} \\
Ours & \textbf{91.13\%} & \textbf{90.06\%} & \textbf{77.10\%} & 3.238 \\
\hline
\multicolumn{5}{c}{\textit{Test on DadaGP (trained on DadaGP + Leduc)}} \\
\hline
Fretting Transformer & 80.02\% & 84.10\% & 71.47\% & \textbf{3.311} \\
Ours & \textbf{91.72\%} & \textbf{90.51\%} & \textbf{77.61\%} & 3.315 \\
\hline
\end{tabular}
}
\vspace{4pt}
\caption{Evaluation results on the DadaGP test set under different training data settings.}
\label{tab:dadagp_train_setting}
\end{table}

\begin{table}[t]
\centering
\renewcommand{\arraystretch}{1.15}
\resizebox{\columnwidth}{!}{
\begin{tabular}{l|cccc}
\hline
\textbf{Method} & \textbf{Token Acc.$\uparrow$} & \textbf{Pitch Acc.$\uparrow$} & \textbf{Tab Acc.$\uparrow$} & \textbf{Difficulty $\downarrow$} \\
\hline
\multicolumn{5}{c}{\textit{Test on Leduc (trained on Leduc only)}} \\
\hline
Fretting Transformer & 18.99\% & 0.49\%  & 0.49\%  & \textbf{1.672} \\
Ours & \textbf{97.80\%} & \textbf{98.04\%} & \textbf{98.06\%} & 2.690 \\
\hline
\multicolumn{5}{c}{\textit{Test on Leduc (trained on DadaGP + Leduc)}} \\
\hline
Fretting Transformer & \textbf{95.69\%} & \textbf{95.97\%} & \textbf{95.97\%} & \textbf{2.681} \\
Ours & 95.15\% & 95.24\% & 95.26\% & 2.689 \\
\hline
\end{tabular}
}
\vspace{4pt}
\caption{Evaluation results on the Leduc test set under different training data settings.}
\label{tab:leduc_train_setting}
\end{table}

\subsection{Error Analysis}

\begin{figure*}[t]
    \centering
    \captionsetup[subfigure]{skip=2pt}
    \setlength{\abovecaptionskip}{2pt}
    \setlength{\belowcaptionskip}{0pt}

    \begin{subfigure}[t]{0.49\textwidth}
        \centering
        \includegraphics[
            width=\linewidth,
            height=0.26\textheight,
            keepaspectratio
        ]{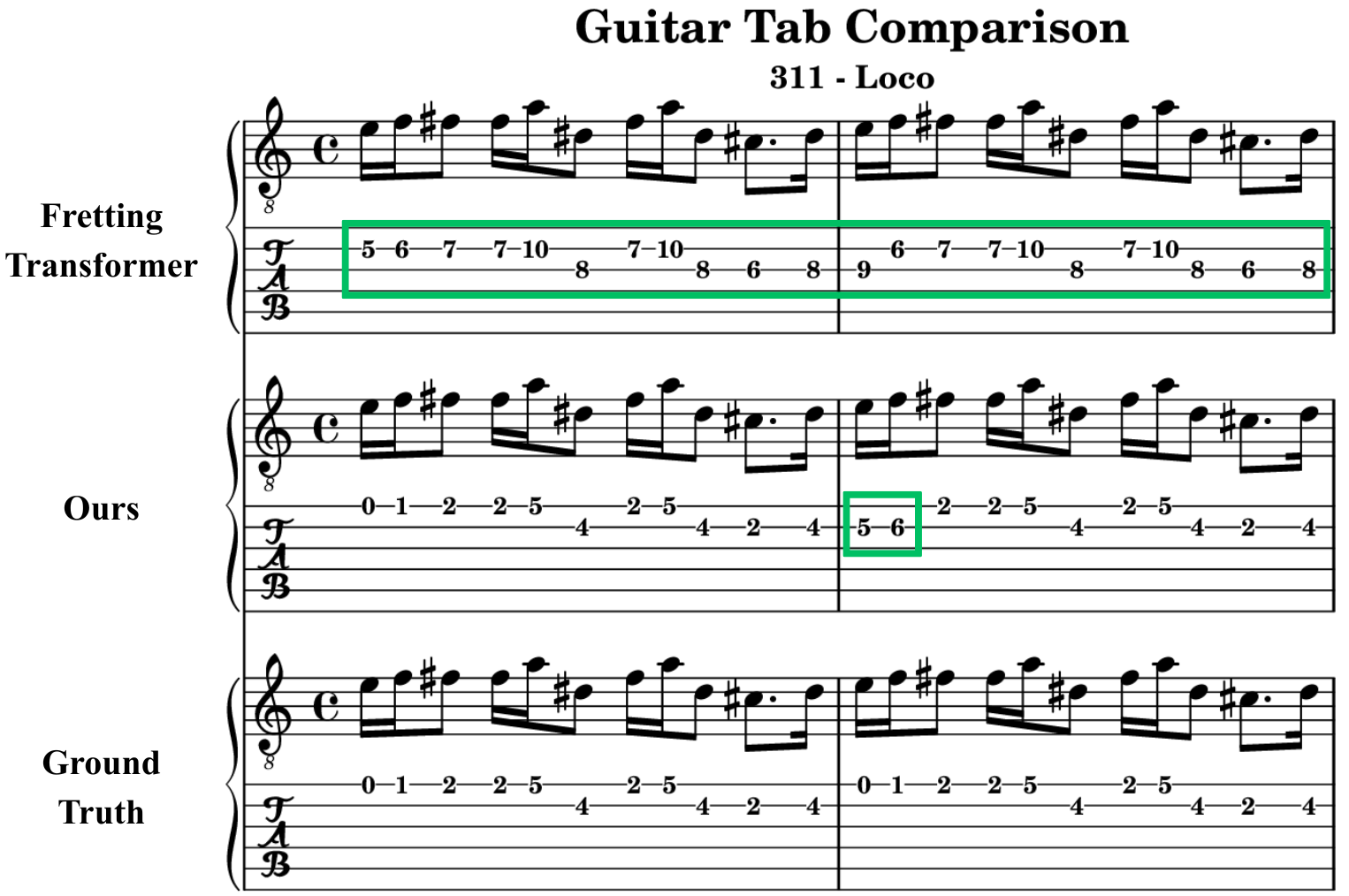}
        \caption{``311 - Loco'' from the DadaGP dataset.}
        \label{fig:guitar_tab_loco}
    \end{subfigure}
    \hfill
    \begin{subfigure}[t]{0.49\textwidth}
        \centering
        \includegraphics[
            width=\linewidth,
            height=0.26\textheight,
            keepaspectratio
        ]{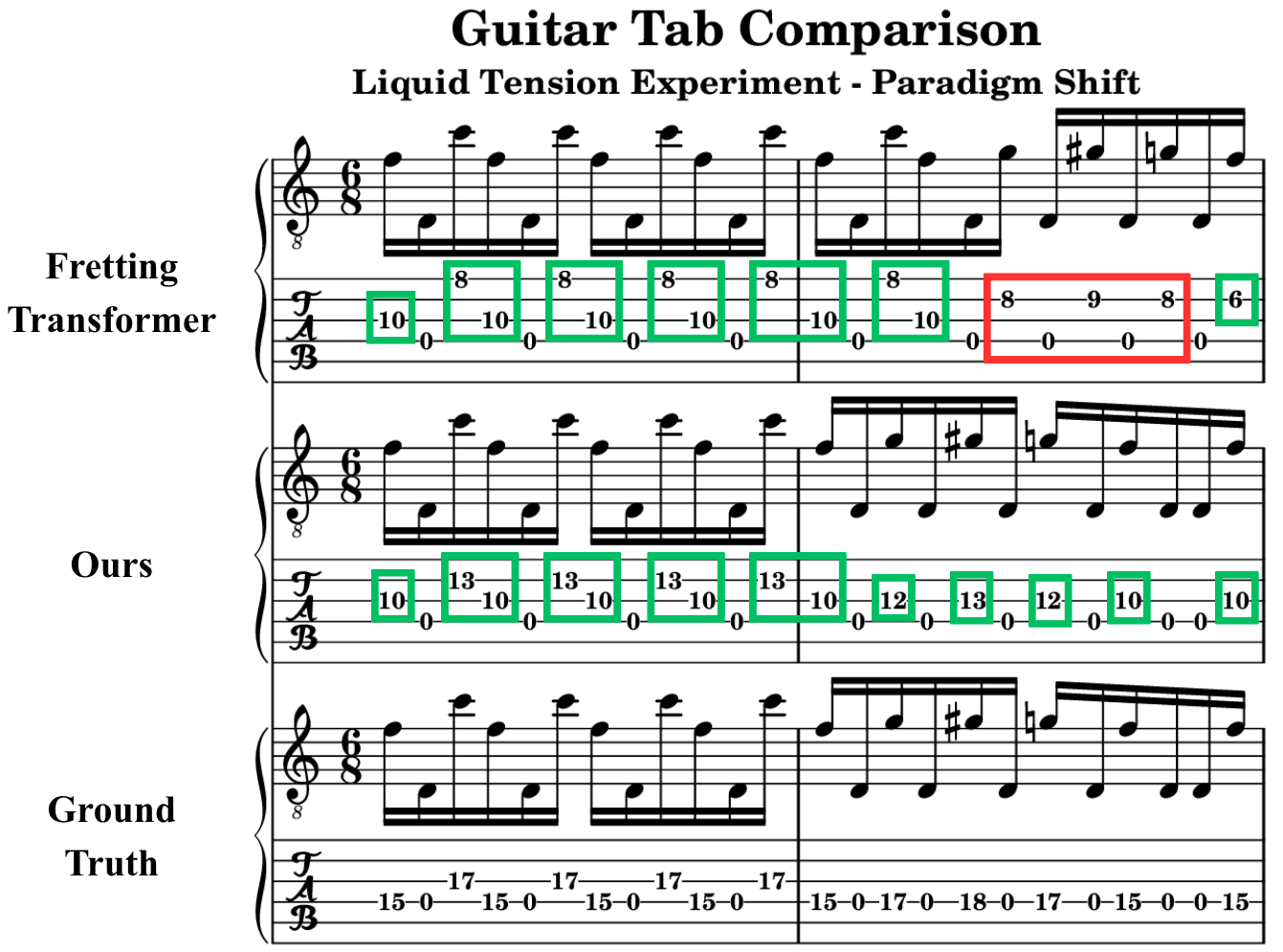}
        \caption{``Paradigm Shift'' from the DadaGP dataset.}
        \label{fig:guitar_tab_paradigm}
    \end{subfigure}

    \vspace{-0.4em}

    \caption{
    Guitar tablature comparison between the Fretting Transformer, our method, and the ground truth. 
    The green boxes highlight position-ambiguity cases, where the predicted notes produce the same pitch as the ground truth but use different string--fret positions. 
    The red boxes highlight pitch errors.
    }
    \label{fig:guitar_tab_comparison}
\end{figure*}

We first analyze the prediction errors under free decoding, where no pitch-validity constraint is applied. To identify the main sources of error, we perform a note-level decomposition and categorize each prediction outcome into timing, pitch, position, or correct classes. 
\TKrevised{
We define the four note-level classes as follows:
\begin{itemize}
    \item \textbf{Timing}: the predicted note event occurs at an incorrect temporal position.
    \item \textbf{Pitch}: the predicted note is temporally aligned with the ground truth, but its pitch is incorrect.
    \item \textbf{Position}: the predicted note has the correct timing and pitch, but its string--fret assignment differs from the ground truth tablature position.
    \item \textbf{Correct}: the predicted  note matches the ground truth in timing, pitch, and string--fret position.
\end{itemize}
}
\TKrevised{
Table~\ref{tab:decoding_error_decomp} reports the performance metrics and the corresponding note-level error decomposition rates on DadaGP, where \textbf{Tab Acc.} corresponds to the correct rate.}
The results show that our method reduces all three types of errors compared with the Fretting Transformer.
In particular, our method produces fewer timing errors, which is consistent with the intended role of explicit \texttt{NOTE\_ON\TKrevised{\_PITCH}} events in providing note-level structure to the decoder. It also reduces pitch errors and position errors, indicating that the proposed framework more often predicts both the correct pitch and the ground-truth string--fret position. Overall, the free decoding error decomposition suggests that our method improves tablature prediction by reducing temporal misalignment, pitch mistakes, and pitch-correct but position-mismatched TAB predictions.

\begin{table*}[t]
\centering
\renewcommand{\arraystretch}{1.2}
\resizebox{0.8\textwidth}{!}{
\begin{tabular}{l|l|cc|ccc}
\hline
\textbf{Method} & \textbf{Constraint} 
& \multicolumn{2}{c|}{\textbf{Performance}} 
& \multicolumn{3}{c}{\textbf{Error Decomposition}} \\
\cline{3-7}
& 
& \textbf{Token Acc.$\uparrow$} 
& \textbf{Tab Acc.$\uparrow$} 
& \textbf{Timing$\downarrow$} 
& \textbf{Pitch$\downarrow$} 
& \textbf{Position$\downarrow$} \\
\hline
\multirow{3}{*}{Fretting Transformer} 
& Free       & 80.57\% & 71.78\% & 4.43\% & 9.91\% & 13.88\% \\
& Rule-based~\cite{fretting-transformer} & 81.17\% & 72.82\% & 4.43\% & 8.15\% & 14.6\% \\
& Constrained decoding & 90.17\% & 83.65\% & \textbf{0.00\%} & \textbf{0.00\%} & 16.35\% \\
\hline
\multirow{3}{*}{Ours} 
& Free      & {91.13\%} & {77.10\%} & 2.56\% & 7.38\% & \textbf{12.96\%} \\
& Rule-based~\cite{fretting-transformer} & {91.57}\% & {77.41}\% & 2.56\% & 6.82\% & 13.22\% \\
& Constrained decoding & \textbf{95.77\%} & \textbf{85.05\%} & \textbf{0.00\%} & \textbf{0.00\%} & 14.95\% \\
\hline
\end{tabular}
}
\vspace{4pt}
\caption{Performance comparison and error decomposition of the Fretting Transformer and our method under free decoding\TKrevised{, rule-based post-processing and} pitch-validity constrained decoding on DadaGP. Error decomposition values indicate the percentage of each error type.}
\label{tab:decoding_error_decomp}
\end{table*}

To further illustrate these error types, we visualize two representative tablature examples in Figure~\ref{fig:guitar_tab_comparison}. In Figure~\ref{fig:guitar_tab_comparison}(a), the Fretting Transformer often assigns notes to higher fret positions on lower strings, whereas the ground truth uses open or lower-fret positions on higher strings. These predictions can still produce the correct pitch, but they differ from the ground-truth tablature because of string--fret position ambiguity. 
In Figure~\ref{fig:guitar_tab_comparison}(b), the Fretting Transformer also produces pitch errors, highlighted by the red boxes. By contrast, our method predicts the correct pitches in these regions, and its remaining mismatches are mostly pitch-valid string--fret alternatives. 

\subsection{Pitch-Validity Constrained Decoding}


The free-decoding error analysis shows that TAB errors can arise from \TKrevised{three} different sources: \TKrevised{timing mismatch}, pitch-invalid predictions and pitch-valid but ground-truth-mismatched string--fret choices. To further separate these \TKrevised{three} sources of error, we evaluate both methods under the pitch-validity constrained decoding defined in Section~\ref{sec:constrained_decoding}.

We compare two decoding conditions: free decoding, where the model generates tokens without any validity constraint, and pitch-validity constrained decoding, where a logit mask restricts each TAB prediction to the pitch-valid set $\mathcal{T}(p)$. 
This setting removes TAB candidates that cannot physically produce the target pitch and therefore allows us to separate pitch-level errors from pitch-valid string--fret selection errors. 
\TKrevised{
In addition, we compare with rule-base post-processing based on the procedure described in \cite{fretting-transformer}.
Since the source code was not released, we reproduce the method according to the paper description.
Each TAB token is corrected to the nearest playable string--fret position whenever its implied pitch differs from the target pitch specified by the input sequence.}
Table~\ref{tab:decoding_error_decomp} reports the results \TKrevised{under Free decoding, pitch-validity constrained decoding, and rule-based post-processing in the DadaGP-only training setting.}

Compared with free decoding, pitch-validity constrained decoding improves \Kevinrevised{tablature accuracy} for both models. The Fretting Transformer improves from 71.78\% to 83.65\%, a gain
of 11.87 percentage points, whereas our method improves from 77.10\% to 85.05\%, a gain of 7.95 percentage points. The larger gain of the Fretting
Transformer suggests that a greater portion of its free-decoding errors comes from pitch-invalid TAB predictions. In contrast, our method already produces
fewer pitch-invalid predictions under free decoding, so it benefits less from the constraint but still achieves higher \Kevinrevised{token accuracy and tablature accuracy} under pitch-validity constrained decoding.
\TKrevised{We also observe that the rule-based post-processing improves slightly compare with free decoding. However, the improvement is limited because the rule-based post-processing cannot align timing mismatches.}

These results extend the previous error analysis. Pitch-validity constrained decoding improves overall \Kevinrevised{tablature accuracy} for both models. However, it does not eliminate all TAB errors. After pitch-level mistakes are controlled, the remaining errors are concentrated in cases where the model selects a pitch-valid string--fret position that differs from the ground-truth annotation. Thus, pitch-validity constrained decoding reveals an error-composition trade-off: it improves pitch consistency, but the remaining difficulty shifts toward exact string--fret selection among multiple pitch-valid candidates.
This also shows that pitch-valid string--fret ambiguity remains a key challenge even when pitch-invalid predictions are removed.

\section{Conclusion}\label{sec:conclusion}

We proposed a guitar tablature transcription framework that combines explicit note-event tokenization, regularized training, and pitch-validity constrained decoding. 
By incorporating \texttt{NOTE\_ON\TKrevised{\_PITCH}} and \texttt{NOTE\_OFF\TKrevised{\_PITCH}} tokens into the decoder output, the proposed tokenization makes note-event structure more explicit during tablature generation. 
Experiments on the DadaGP and Fran\c{c}ois Leduc datasets show that our framework improves pitch and tablature accuracy over the retrained Fretting Transformer baseline, particularly under the small-scale Leduc training setting.

Our note-level error analysis and pitch-validity constrained decoding experiments show that removing pitch-invalid TAB candidates improves \Kevinrevised{tablature accuracy} for both models, but exact string--fret selection remains challenging even when pitch-level errors are controlled. In future work, we plan to incorporate fingering information and playability-aware inference to better model performer-preferred tablature choices.



\bibliographystyle{IEEEtran}
\bibliography{references-new}

\end{document}